# On the characterization of a random monolayer of particles from coherent optical reflectance

**F. Alarcón-Oseguera,[1] M. Peña-Gomar,[1*] A. García-Valenzuela,[2] F. Castillo,[3] and E. Pérez[3]**

[1] *Facultad de Ciencias Físico-Matemáticas de la Universidad Michoacana de San Nicolás de Hidalgo, Ciudad Universitaria,58030, Morelia, Michoacán, México*

[2] *Centro de Ciencias Aplicadas y Desarrollo Tecnológico, Universidad Nacional Autónoma de México, Cuidad Universitaria, A.P. 70-168,04510, México D.F.*

[3] *Instituto de Física, Universidad Autónoma de San Luis Potosí, Alvaro Obregón 64, C.P. 78000, San Luis Potosí, SLP, México*

**Corresponding author:* mgomar@fismat.umich.mx

We present the viability of obtaining the particle size and surface coverage in a monolayer of polystyrene particles adsorbed on a glass surface from optical coherent-reflectance data around the critical angle in an internal reflection configuration.

The study of particle adsorption onto surfaces is relevant to materials science and colloid science [1]. The particle adsorption process in a liquid-solid interface cannot be described by a general theory, and is proving to be a quite complex problem [2]. Reliable and sensitive experimental methods designed to monitor and characterize monolayer formation are then essential in understanding the physics of this problem. Optical methods for the study of the particle adsorption in flat surfaces have been used for some years, showing a good performance. Two examples of these methods are: reflectometry [3] and elipsometry [4]. When particles are not very small compared to the wavelength of the light, the reflected light coming from a monolayer of particles can not be described by an effective medium theory, and a model based on the coherent-scattering reflectance represents an available approach.

Measuring the reflectivity of a monolayer adsorbed on a flat glass-water surface around the Brewster's angle has already been used for characterizing random monolayers of particles [5,6]. In those works, a simple coherent scattering model (RCM) was proposed for monolayers with low surface coverage and TM-polarized light incident around Brewster's angle, and it was fitted to the data determining the particle radius and surface coverage with acceptable precision. Recently a new and more robust CSM for coherent reflectance from monolayers with low surface coverage has been proposed, which is valid for all angles of incidence, including around and behind the critical angle in an internal reflection configuration [7]. In that work, the model is compared with experimental reflectance around the critical angle and found that the model could reproduce the experimental data within the experimental uncertainties.

In practice, it is easier to measure precisely optical-reflectivity curves around the critical angle than around Brewster's angle. The reason is that the values of the reflectivity of TM polarized light around Brewster's angle are very small and the spurious signals due to imperfections of the experimental system, and stray light may introduce large errors which does not occur around the critical angle experiments. Therefore the question arises if it is also possible to determine particle size and surface coverage form reflectivity curves around the critical angle in an internal reflection configuration of a glass interface supporting a random monolayer of particles. This is the main question we address in this letter.

This coherent scattering model (CSM) describes the coherent reflection of light from a monodisperse monolayer of particles randomly adsorbed on a flat surface. The model takes into account multiple scattering effects among the particles and with the interfase and can be used with large particles and for any angle of incidence, but it is limited to small surface coverage. Herein we will assume that the incident beam is a laser beam with a Gaussian intensity profile. The formula to calculate the reflectivity in this case is given in details in Ref. [7].

To evaluate numerically the CSM given in Ref. [7] it is necessary to calculate the elements of the amplitude-scattering matrix of an isolated particle. It is assumed that all particles have the same radius. For simplicity in this work, we consider only cases where the refractive index contrast between the particles of the monolayer and the surrounding medium is small and particles are not too large; and thus, we use the Rayleigh-Gans approximation [8] to calculate the elements of the amplitude-scattering matrix of the particles.

In order to test the viability of retrieving the particle radius, $a$, and surface coverage, $\Theta$, from coherent reflectance curves around the critical angle in an internal reflection configuration, we used simulated data as well as actual experimental data. The usual route to extract the values of $a$ and $\Theta$ from a reflectivity versus angle-of-incidence curve is to fit the CSM to the experimental data by adjusting the values of $a$ and $\Theta$, and assuming all other parameters known and fixed.

The methodology for parameter fitting consists of finding the minimum of the function $\chi^2$ defined as:

$$\chi^2 = \frac{1}{\nu}\sum_{i=1}^{N}\left(\frac{R_e(\theta_i) - R_m(\theta_i; a, \Theta)}{\sigma_I(\theta_i)}\right)^2, \qquad (1)$$

where $\nu = N\text{-}M$ is the degrees of freedom on the parameter fitting, $N$ and $M$ are the number of experimental data and the number of parameters to fit, respectively; $\sigma_I$ is the standard deviation of the error of the experimental reflectivity, $R_e(\theta_i)$, which is assumed constant for all the data, and $R_m(\theta_i; a, \Theta)$ is the reflectivity calculated with the coherent model. To minimize the function $\chi^2$ in the $(a,\Theta)$ space we implemented a computational program that searches the zero of the derivative of $\chi^2$ respect to fitted parameters, the iterative method is based on the Newton-Rapson method and on an inversion by singular values of the matrix formed by these derivatives [9].

We simulated experimental data by calculating the coherent reflectance data with the reflectance model reported in Ref. [8] and adding gaussian noise to it. Specifically, the simulated data is of the form,

$$R_{sim}(\theta_i, x) = [1 + sg(x)]R_m(\theta_i), \qquad (2)$$

where $g(x)$ is a gaussian random function with zero mean value and standard deviation equal to one. The variance of the simulated intensity is then given by $\sigma_I^2 = s^2 R_m^2(\theta_i)$.

We obtained simulated data for TM reflectivity in an internal reflection configuration assuming a relative error of $s = 0.01\%$ and $s = 1\%$ for different values of $a$ and $\Theta$. In all cases we assumed particles of refractive index $n_p$ = 1.59 (polyestirene) adsorbed on a glass–water interfase. The refractive index of the glass and water were supposed to be 1.515 and 1.331, respectively. In all examples we chose a wavelength of light of 632nm.

Figures 1 and 2 are examples, respectively, of a simulated and a experimental fitting practiced to reflectance data. The first case corresponds to $s$ = 0.01%, $a$ = 150 nm and $\Theta$ = 3%, while the second one correspond to particles of 149 nm with a $\Theta$ = 8%, approximatively. In these figures, the full circles are the simulated or experimental data (0.5 degrees apart). Additionally, in the insets of Figs. 1 and 2 we show contours in the ($a$, $\Theta$) space of constant values of $\chi^2$ for each example to help us asses more clearly the viability of obtaining $a$ and $\Theta$. The height and width of the region in the ($a$, $\Theta$) space enclosed by the contours of constant $\chi^2$ give us a measure of the possible uncertainty in recovering $a$ and $\Theta$. We found that fitting the CSM model to the simulated data resulted in values of $\chi^2$ less than 1. Therefore, we defined the uncertainty in recovering $a$ and $\Theta$ in these examples as the height and width of the corresponding contour for $\chi^2$ = 1. The results for all examples studied here are summarized in table 1. In the first column we give the nominal values for $a$ and $\Theta$, in the second and fourth columns we give the values retrieved from the fitting procedure for $a$ and $\Theta$ respectively, and in the third and fifth columns we give the corresponding uncertainties estimated from the contours of $\chi^2$ = 1.

Additionally, we fitted our CSM to actual experimental data reported previously in Ref. [8]. We considered two cases: one with nominal values of $a$ = 149 nm and $\Theta \approx 8$ and the other for = 256 nm and $\Theta \approx 3$ %. The values of the particle radius were determined with dynamic light scattering and the values of the cover fraction were roughly estimated from images by optical microscopy. The relative error on the reflectivity measurements was about 1% for all experimental points (estimated from noise at the output of the photodetector). The CSM curves fitted to the experimental data reached a minimum value of $\chi^2$ of 1.99. The retrieved values of *a* and Θ as well as the uncertainties estimated from the contours of $\chi^2 = 1$. These examples are also given in table 1.

It is clear from table 1 that the relative uncertainties in retrieving the values of *a* and Θ increases as the relative error of the measured reflectivity values, *s*, increases, but not in the same proportion. For instance, when the value of *s* changes by a factor of 100, from $10^{-4}$ to $10^{-2}$, the uncertainties on *a* and Θ grow about a factor of 10.

Note that for the example with $s = 1\%$ and a particle radius of $a = 50$ nm we could not calculate an uncertainty. The reason is that, in this case, a continuum of pairs of values of *a* and Θ could fit well the CSM to the simulated data. This is because the contours of constant values of $\chi^2$ were not closed showing that there is a limit in the smallest size of particles which can be measured by this method. This can actually be explained from the fact that as the particle size reduces, each individual particle scatters light more isotropically, and at some point, the scattering pattern of an isolated particle does not depend on the particle radius. The scattering efficiency does depend on particle's volume, but also on their number density. Therefore, it becomes indistinguishable if the cover fraction or the particle's radius change.

For larger particles, we believe the achievable uncertainties reported in table 1 are actually very promising and suggest that retrieving *a* and Θ from reflectivity curves around the critical angle can become a practical method for characterizing colloidal particles adsorbed on a flat surface, as long as these are not too small.

To compare our results summarized in table 1 with those obtained in the reference [6], we also simulated data for TM reflectivity around Brewster's angle assuming a value for the standard deviation of the reflectivity of, $s = 1\%$. We plotted the contours for constant values of the function $\chi^2$ and found that these contours are similar in size with those obtained for simulated data of TM reflectivity curves around the critical angle with the same standard deviation. However, experimentally, to measure the reflectivity around the Brewster's angle with a relative error of 0.01 requires additional considerations and is in general more difficult than to measure reflectivity curves with the same 0.01 relative error around the critical angle.

In conclusion, we have found that fitting a CSM to optical reflectivity curves in an internal reflection configuration around the critical angle with a random monolayer of particles adsorbed on the surface can in fact provide the particle's radius and surface coverage once the particles are sufficiently large.

It is generally easier to measure reflectivity curves around the critical angle than it is to measure them around Brewster's angle with similar relative errors. Therefore, in practice, it appears more advantageous to work with coherent reflectance curves around the critical angle than around Brewster's angle. It will be interesting in the future to extend the validity of the CSM to higher values of the surface coverage.

**Acknowledgments**

We are grateful to Dirección General de Asuntos del Personal Académico from Universidad Nacional Autónoma de México and Consejo Nacional de Ciencia y Tecnología (México) for financial support through grants PAPIIT IN-116106 and CB-05-01 No. D49482-F respectively.

## Figure captions

***Figure 1.*** *Simulated data of TM reflectance around critical angle from a monolayer of particles with a = 150 nm and* Θ *= 3% (circles) with* $\sigma_I = 1x10^{-4}$*. The fitting (solid curve) produces a =149.32 nm and* Θ *= 3.018% with* $\chi^2$*=0.8478,* $\sigma_\Theta$*=3.028% and* $\sigma_a$*=0.12 nm. The dotted line and dash-dote lines are successive fitting to these data. The contour of* $\chi^2$ *in the parameter space is shown in the inset of the Figure.*

***Figure 2.*** *Experimental data of TM reflectance around the critical angle from a monolayer of particles with nominal values of a = 149 nm and Θ = 8%, approximately (circle). The lines have the same meaning that Figure 1 and the fitting produces, a =134.9 nm and Θ = 8.99 % with* $\chi^2$ *= 1.9942,* $\sigma_\Theta$*=0.31% and* $\sigma_a$*=4.1. The experimental error was of 1%.*

**Figure 1**

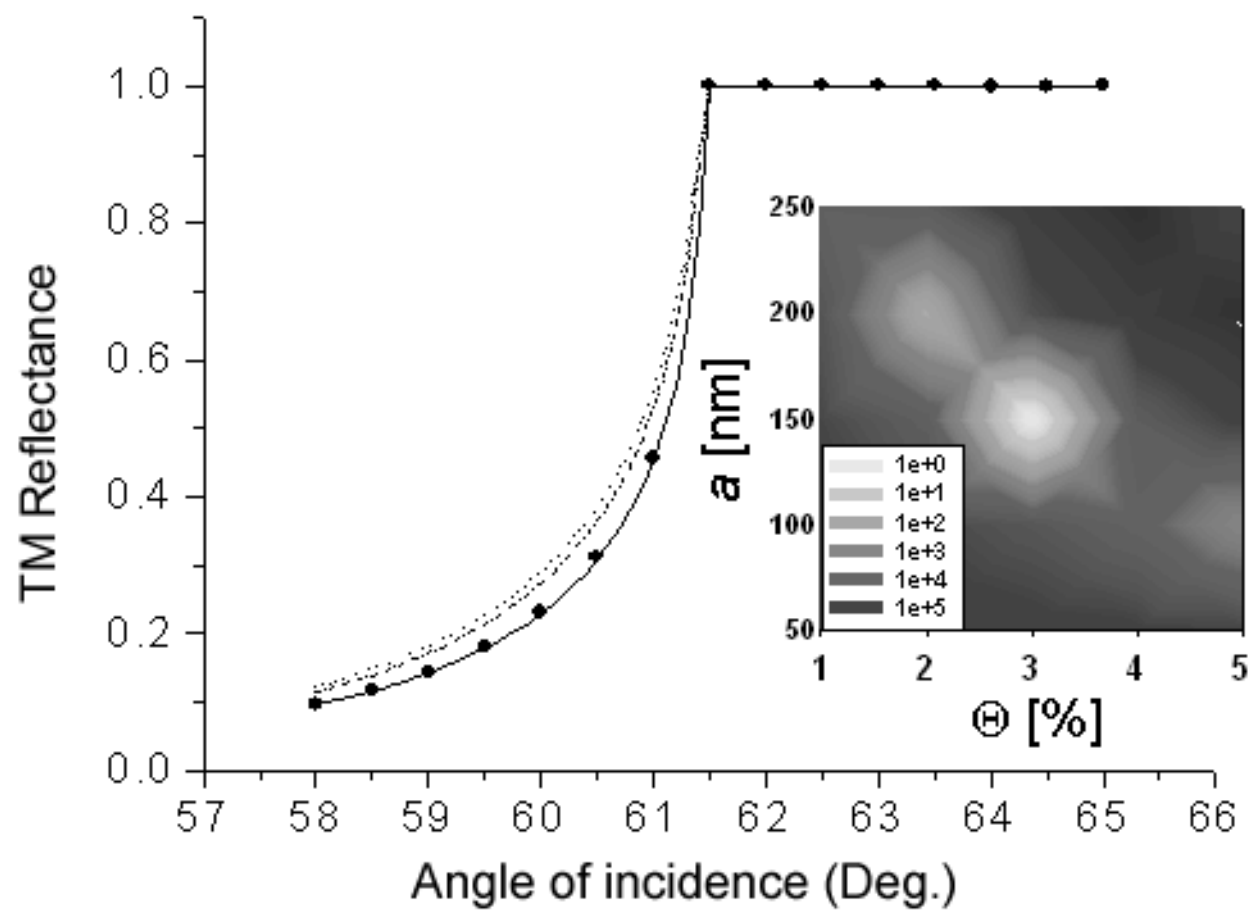

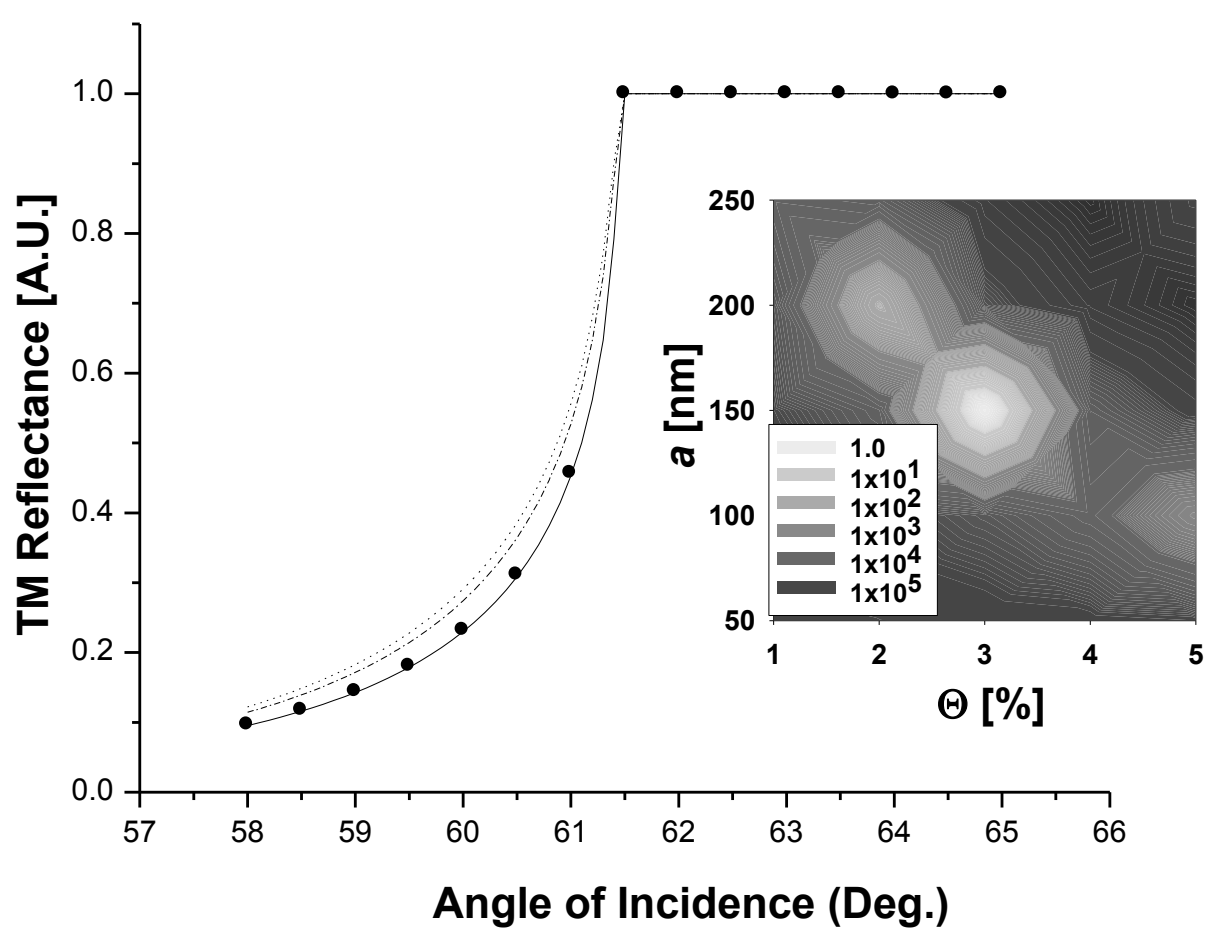

**Figure 2**

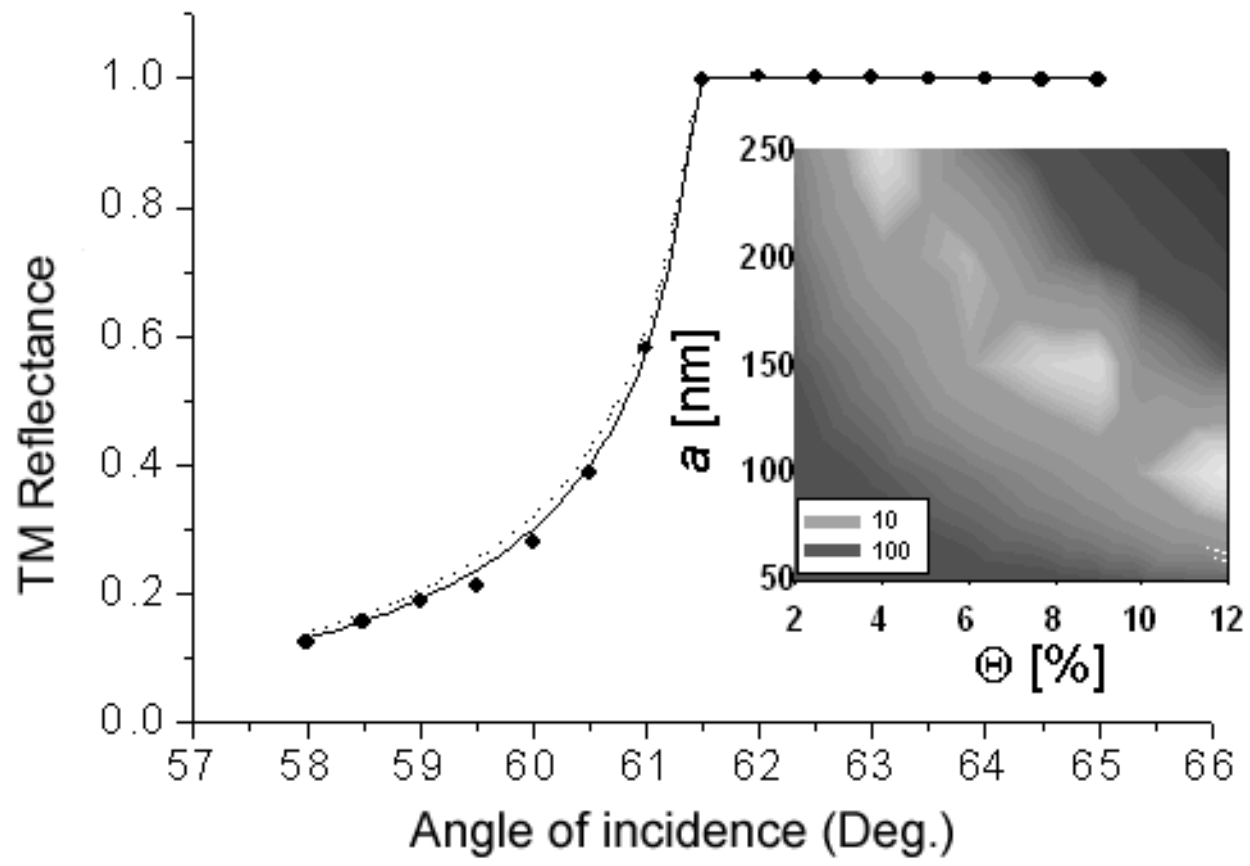

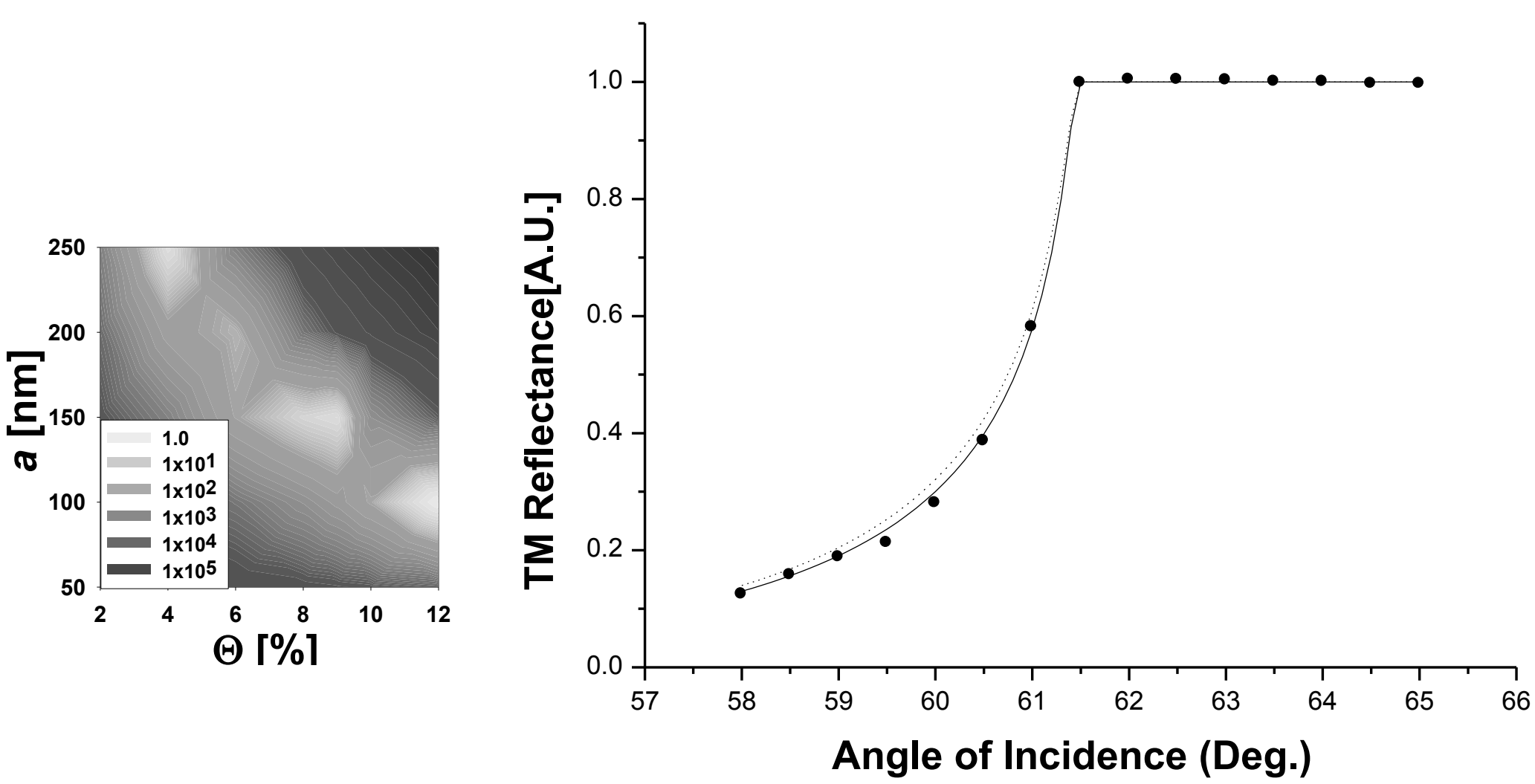

**Table 1.** Simulated and experimental data.

| | σ=$10^{-4}$ | | | |
|---|---|---|---|---|
| **Simulated Parameters [*a*,Θ]** | ***a*(nm)** | **σ *a* (nm)** | **Θ(%)** | **σ Θ(%)** |
| 50nm, 3% | 49.87 | ±0.24 | 3.033 | ±0,025 |
| 150nm, 3% | 149.32 | ±0.12 | 3.018 | ±3.4x$10^{-3}$ |
| 250nm, 3% | 250.00 | ±0.37 | 2.999 | ±1.9x$10^{-3}$ |
| 350nm, 3% | 350.00 | ±2.0 | 2.999 | ±2.1x$10^{-3}$ |
| 250nm, 1% | 249.60 | ±1.0 | 1.0012 | ±2.0x$10^{-3}$ |
| 250nm, 8% | 250.00 | ±0.2 | 7.998 | ±0,019 |
| | | | | |
| | **σ=10^-2** | | | |
| **Simulated Parameters [*a*,Θ]** | ***a*(nm)** | **σ *a* (nm)** | **Θ(%)** | **σ Θ(%)** |
| 50nm, 3% | 41 | NA | 3.52 | NA |
| 150nm, 3% | 149 | ±3,25 | 2.87 | ±0.09 |
| 250nm, 3% | 251 | ±2,25 | 2.89 | ±0.03 |
| 350nm, 3% | 351 | ±2,38 | 2.9 | ±0.02 |
| 250nm, 1% | 248 | ±16 | 0.93 | ±0.06 |
| 250nm, 8% | 250 | ±11,25 | 7.86 | ±0.3 |
| | | | | |
| | | | | |
| | **Experimental error of 5%** | | | |
| **Experimental Parameters** | ***a*(nm)** | **σ *a* (nm)** | **Θ(%)** | **σ Θ(%)** |
| Nominal values | 134.9 | ±6,25 | 8.99 | ±0.64 |
| Nominal values | 255 | ±3,75 | 3.08 | ±0.035 |